\documentclass[a4paper,aps,floatfix,twocolumn,prb]{revtex4}
\usepackage[latin1]{inputenc}
\usepackage[OT1]{fontenc}
\usepackage{amsmath}
\usepackage{amssymb,amsfonts}
\usepackage{pifont}
\usepackage{graphicx}
\usepackage{color}

\newcommand{\Eqref}[1]{Eq.~\eqref{#1}}

\newcommand{\vect}[1]{\mathbf{#1}}

\newcommand{\n}{\vect{n}}
\newcommand{\p}{\vect{p}}
\newcommand{\q}{\vect{q}}

\newcommand{\e}{\mathrm{e}}
\newcommand{\E}{\vect{E}}
\newcommand{\G}{\vect{G}}
\newcommand{\bE}{\bar{\E}}

\renewcommand{\P}{\vect{P}}
\newcommand{\R}{\vect{r}}
\newcommand{\eps}{\epsilon_0}

\newcommand{\DD}{\mathcal{D}}
\newcommand{\Z}{{Z}}

\newcommand{\Zfl}{\Z_\text{fluct}}
\newcommand{\Zcoul}{\Z_\text{Coulomb}}

\newcommand{\Pe}{\phi^\text{(ext)}}
\newcommand{\lo}{_\parallel}

\def\D{\mathrm{d}}
\newcommand{\DR}{\ensuremath{\:\D³r}}
\newcommand{\DRj}{\ensuremath{\prod_j\DR_j}}
\newcommand{\grad}{\nabla}
\renewcommand{\div}{\nabla \cdot}
\newcommand{\curl}{\nabla \times}

\begin{document}

\title{Boundary conditions in local electrostatics algorithms}
\author{L.  Levrel}
\affiliation{Physique des Liquides et Milieux Complexes, Facult\'e
des Sciences et Technologie, Universit\'e Paris Est (Cr\'eteil), 61 avenue du
G\'en\'eral-de-Gaulle, F-94010 Cr\'eteil Cedex, France}
\affiliation{IUFM de l'Acad\'emie de Cr\'eteil, Rue Jean Mac\'e, F-94861 Bonneuil-sur-Marne
Cedex, France}
\author{A. C. Maggs}
\affiliation{Physico-Chimie Th\'eorique, Gulliver-CNRS, ESPCI,
10 rue Vauquelin, Paris  75005, France.}
\date{\today}
\begin{abstract}
  We study the simulation of charged systems in the presence of general
  boundary conditions in a local Monte Carlo algorithm based on a constrained
  electric field.  We firstly show how to implement constant-potential,
  Dirichlet, boundary conditions by introducing extra Monte Carlo moves to the
  algorithm.  Secondly, we show the interest of the algorithm for studying
  systems which require anisotropic electrostatic boundary conditions for
  simulating planar geometries such as membranes.
\end{abstract}
\maketitle
\section{Introduction}

In the simulation of condensed matter one very often imposes periodic boundary
conditions in order to minimize surface artefacts which give rise to slow
convergence of energies to their bulk values. While such boundary conditions
are simple to understand for short-ranged potentials they lead to many
subtleties when working with charged systems \cite{deleeuw,fraser}.  Attempts
at simplifying the problem by using truncated potentials lead to violations of
basic sum rules on the structure factor \cite{stillinger}, incorrect number
fluctuations in finite systems \cite{lebowitz} or even loss of electrical
conductivity \cite{marchetti}.  The correct treatment of charged media
requires a full treatment of the long-ranged Coulomb interaction.

Classical methods of treating the Coulomb interaction use careful mathematical
analysis to convert the conditionally convergent Coulomb sum into a well
defined mathematical object such as the Ewald sum \cite{deleeuw}.  Recently we
introduced an alternative treatment of the Coulomb problem that allows one to
replace the \textsl{global}\/ calculation of the interaction energy by a local
dynamic process \cite{PRL88, JCP117, StatPhys}. The main interest in this
transformation is that it enables one to perform a local Monte Carlo
simulation in the presence of charges and dielectrics, without ever solving
the Poisson equation.  The algorithm works by evolving the electric field in
time (in a manner similar to Maxwell's equations) while eliminating
``uninteresting'' degrees of freedom such as the magnetic field.  The
algorithm uses the energy
\begin{equation}
  U =   \int \frac {\epsilon_0 \E^2}{2}  \DR \label{energy}
\end{equation}
where $\E$ is the electric field while implementing Gauss' law
\begin{equation}
  \epsilon_0 \div \E - \rho=0 \label{constraint}
\end{equation}
as a dynamic constraint.  We showed \cite{JCP120} that the algorithm can be
used to generate \textsl{tin-foil}\/ or \textsl{vacuum}\/ boundary conditions if one
chooses appropriate dynamics for the $\q=0$ component of the electric field.
Until now applications have been to the properties of bulk, three dimensional
media \cite{igor,PRL93} in periodic systems.

This paper generalizes the method to a broader class of physical systems and
boundary conditions.  Firstly we consider the case, particularly important in
devices and electrodes, of imposition of an external potential on a metallic
surface; a case which requires the use of Dirichlet boundary conditions for
the potential. Again we find that the locality of the formulation allows the
simulation of a broader class of geometries, including those for which the use
of fast Fourier techniques is difficult.

A second generalization of the method is required to treat \textsl{anisotropic}\/
systems, in particular membranes. The simulation of thin, quasi
two-dimensional systems in three-dimensional space is surprisingly difficult
\cite{berkoslab, holm53, holmslab}.
The use of a purely local algorithm requires only small modifications in order
to minimize finite size effects; we argue that it will also give rather
favorable complexity in the limit of large number of particles, $N$.

Our paper contains two, independent, self-contained theoretical
sections. Firstly in II we treat the problem of metallic boundary conditions
including certain subtleties as to how a true metal behaves. Secondly in III
we consider electrostatics in anisotropic $2+1$ dimensional geometries.
Implementation of the ideas has been performed by Thompson and Rottler
\cite{comp} using off lattice techniques suitable for atomistic simulation.
In their paper they compare detailed simulations with available theories.

\section{Metallic boundary conditions}

We begin by emphasizing the difference between idealized metallic boundaries
and true physical systems in which screening occurs over a small but finite
Debye length. We then show how to impose Dirichlet conditions on the
potential, $\phi$ while still keeping $\E$ as the main dynamic variable.

\subsection*{The nature of metallic boundary conditions}

There are two different, self-consistent ways of calculating the energy of
charges in the presence of a dielectric or conducting interfaces. In the first
treatment we calculate the potential energy arising from the solution of the
Poisson equation
\[
\div (\epsilon \grad \phi) = - \rho
\]
with $\rho$ the density of free charges, with the appropriate boundary
conditions for each configuration.  The boundary condition at a metallic
surface is that the tangential electric field vanishes, while the normal field
at the surface satisfies
\begin{equation} \vect{D} \cdot \n =\epsilon \E \cdot \n=-\sigma \label{bc}
\end{equation}
where $\sigma$ is the surface charge density. The field is identically zero
within a conductor \cite{landau}. The internal energy is then
\begin{equation}
  U = \frac{1}{2}\int \phi \rho \DR. \label{U0}
\end{equation}
In this case the longitudinal electric field is static, and is given by $\E =
-\grad \phi$. We note that in this treatment thermal fluctuations play no role
so that certain fluctuation phenomena such as thermal Casimir interactions are
neglected.

A physical system at a non-zero temperature always presents polarization or
charge fluctuations; superposed on the interaction energy \Eqref{U0}
is an effective potential coming from these fluctuations.  We now work in the
Debye--H{\"u}ckel limit to calculate an approximate correlation function for
the electric field in a conducting system at finite temperatures. This
illustrative calculation shows that transverse field fluctuations become
important as soon as the temperature $T>0$.

We consider a one component plasma, with neutralizing background. We
approximate the free energy of a conductor as a sum of the electric field
energy \Eqref{energy} and the configurational entropy of the charges with
the functional \cite{JCP120}
\begin{equation}
  F = \int \left \{ 
    \frac{\epsilon_0 \E^2 }{2} + k_BT \; c\ln{c}
  \right \}\DR
  \label{debye}
\end{equation}
where $c(\vect{r})=\rho(\vect{r})/e$ is the number density of mobile ions and
$e$ the charge of the particles. In the limit of small fluctuations of density
$\delta c(\vect{r})$ we expand to second order and use charge conservation so
that $\int \delta c \DR=0 $. Then
\[
F(\delta c, \E) = \int \left \{ \frac{\epsilon_0 \E^2 }{2} + k_BT \; \frac
  {( \delta c)^2 }{2 c_0} \right \}\DR
\]
where $c_0$ is the mean background density of the ions. We now eliminate the
charge fluctuation with the help of Gauss' law and find an effective action
for the electric field
\begin{equation}
  F_E= \int 
  \epsilon_0 \left \{
    \frac {\E^2 }{2} + \frac{(\div \E)^2 }{2\kappa^2}
  \right \} \DR
  \label{fieldenergy}
\end{equation}
with $\kappa$ the inverse Debye length. From \Eqref{fieldenergy} we read
off the longitudinal and transverse correlations of the electric field,
\begin{align*}
  \epsilon_0 \langle \E(q) \E(q) \rangle_{long}
  &= \frac {k_B T }{ 1 + q^2/\kappa^2},  \\
  \epsilon_0 \langle \E(q) \E(q) \rangle_{tran}&=k_B T.
\end{align*}
On wavelengths large compared with the screening length ($q/\kappa <<1$) all
longitudinal and transverse electric modes fluctuate with the same amplitude.

It is instructive to compare with similar calculations for an explicit model of
dielectric in terms of polarization fluctuations \cite{polardiel}. We find
that
\begin{align*}
  \epsilon_0 \langle \E(q) \E(q) \rangle_{long}
  &= k_B T \; \frac{\epsilon -\epsilon_0 }{ \epsilon },  \\
  \epsilon_0 \langle \E(q) \E(q) \rangle_{tran}&=k_B T.
\end{align*}
Again transverse field correlations are unchanged in the presence of
a dielectric or classical charged fluid, longitudinal fluctuations depend on
the material properties. In the limit $\epsilon \rightarrow \infty$ the
fluctuations of a dielectric
are the same as those of a metal for $q \ll\kappa$.

From these expressions in Fourier space we calculate the correlations of the
electric field in real space. For a dielectric (with $\epsilon < \infty$) we
find that the field displays dipolar correlations so that
\[
\langle E_i(\vect{r}) E_j(\vect{r}') \rangle \sim \frac {1}{ |\vect{r}-\vect{r}'|^3 },
\] field correlations are long-ranged. This dipolar correlation in the fields
leads to long-ranged Casimir/Lifshitz interactions between dielectrics.  For a
conducting system field correlations decay exponentially with characteristic
length the Debye length,
\[
\langle E_i(\vect{r}) E_j(\vect{r}') \rangle \sim e^{-\kappa |\vect{r}-\vect{r}'|}.
\]

From these considerations we see that different strategies are needed to
simulate the idealized classical metallic boundary condition \Eqref{bc} or a
true fluctuating charged fluid.  In the first situation one only needs
information on the fields in the nonconducting regions.  In the second case
the surface field is influenced by fluctuations that occur within a few Debye
lengths of the surface. The surface must then be simulated explicitly with an
expression such as \Eqref{debye} in order to obtain results including
Casimir/Lifshitz type interactions. We note that such interactions are very
weak on the macroscopic scale. Between two plates of area $A$ separated by a
distance $H$ the thermal interactions are given by \cite {kardar}
\[
{U}=-A k_BT \frac{\zeta(3) }{ 16 \pi H^2}
\]
with $\zeta(3) \approx 1.20$

In this paper we will only consider the case of imposing the first type of
metallic boundary conditions in which fluctuation forces are neglected.

\subsection*{Fixed-potential boundary conditions}

\subsubsection*{Minimization principle}

We now apply our local simulation algorithm to fixed-potential boundary
conditions of the type \Eqref{bc}. The method of attack is a generalization of
the methods of Ref.~\onlinecite{PRL88}. Firstly we seek a variational principle for the
electric field for which the energy is a \textsl{true minimum}\/ for the geometry
of interest.  We then promote the constraints of the minimization principle to
$\delta$-functions in a partition function.  This partition function is then
shown to generate the same relative statistical weights as the original
variational energy, but is much more convenient for the purposes of simulation
since it allows the local update of fields in a Monte Carlo algorithm.

As noted in the introduction electrostatic interaction is calculated by
minimizing the electric field energy \Eqref{energy} in the presence of the
constraint of Gauss' law, \Eqref{constraint}, using the functional
\begin{equation*}
  A[\E]=\int\left[\frac{\eps \E^2}{2}+\lambda(\eps\div\E-\rho)\right]\DR,
\end{equation*}
where $\lambda$ is a Lagrange multiplier. We now generalize Ref.~\onlinecite{schwinger}
and consider a system with both free charges and conductors, $i$. These
conductors are maintained by external sources at fixed potentials $\Pe_i$.
This is most conveniently done by performing Legendre transform \cite{landau}
in order to eliminate for the unknown surface charge densities $\sigma
=\{\sigma_i\}$ on the surfaces $S=\{S_i\}$ in order to replace them by the
potentials $\Pe=\{\Pe_i\}$:
\begin{equation}\label{foncpotfix}
  U_2[\E,\sigma]= U[\E]- \oint \sigma\Pe \D S.
\end{equation}
We now impose the boundary condition \Eqref{bc} with a further Lagrange
multiplier $\mu$ living on each surface and consider the functional
\begin{equation}
  A_2 = U_2 + \int \lambda(\eps\div\E-\rho)\DR
  + \oint \mu(\eps\n\cdot\E+\sigma)\;\D S. \label{a2}
\end{equation}
In order to find the stationary point of $A_2$ we vary all the fields,
and integrate by parts using $\lambda\div\delta\E=\div(\lambda\delta\E)
-(\grad\lambda)\cdot\delta\E$ and find
\begin{multline*}
  \delta A_2=\int[\eps\;\delta\E\cdot(\E-\grad\lambda)
  +\delta\lambda\;(\eps\div\E-\rho)]\DR\\
  -
  \oint[\delta\sigma(\Pe+\mu)
  +\delta\mu(\eps\n\cdot\E+\sigma)
  -\eps(\lambda-\mu)\delta\E\cdot\n]\;\D S.
\end{multline*}
At the stationary point we find
\begin{alignat*}{2}
  \delta\E &:\quad \left\{\begin{aligned}
      &\E-\grad\lambda=0\\
      &\lambda-\mu=0 \end{aligned}\right.
  &\quad&\begin{aligned} & \text{in $V$,}\\
    &\text{on  $S$,} \end{aligned} \\
  \delta\sigma &:\quad \Pe +\mu =0 &&\text{\,on  $S$,}  \\
  \delta\lambda &: \quad \eps\div\E -\rho =0 &&\text{\,in $V$,} \\
  \delta\mu &: \quad \eps\n\cdot\E +\sigma =0 &&\text{\,on $S$.}
\end{alignat*}

The first equation implies that $\E$ is the gradient of a potential,
$\phi=-\lambda$. The next two imply that $\phi=\Pe$ on $S$ so that we
indeed generate Dirichlet conditions.  Finally the last two equations impose
Gauss' law.  At the stationary point thus we have
$\E=\E_p=-\grad\phi_p$ and $\sigma=\sigma_p$, where the
index $p$ signals the solution to Poisson's equation with Dirichlet boundary
conditions.

\subsubsection*{Partition function}
We now generalize the stationary principle to finite temperatures and replace
the Lagrange multipliers of \Eqref{a2} by $\delta$-functions in a
functional integral.  Let us define a partial partition function,
where the bulk charge distribution $\rho$ is given, as
\begin{equation}
  Z_\rho=\int\DD\E\ \DD\sigma\;
  \delta(\eps\div\E-\rho)\:\delta(\eps\n\cdot\E+\sigma)
  \:\mathrm{e}^{-\beta U_2[\E,\sigma]}.
  \label{zrho}
\end{equation}
We integrate over the electric field, constrained by Gauss' law and also
all values of the surface charge $\sigma$ compatible with the flux
condition at the surface of the conductors.

The full partition function is calculated by integrating over the position of
all mobile charges.  In order to find the interaction in terms of the minimum
of the functional it is convenient to change integration variables so that
$\E=\E_p+\vect{e}$ and $\sigma=\sigma_p+s$.  After noting that
$\sigma_p=-\epsilon_0 \n\cdot\E_p$ we find
\begin{align*}
  Z_\rho
  &=\int\DD\vect{e}\ \DD s\;\delta(\div\vect{e})\:\delta(\eps\n\cdot\vect{e}+s)\\
  &\times \qquad \e^{-\beta\left[\int \frac{\eps}{2}(\E_p+\vect{e})^2\DR
      -\oint (\sigma_p+s)\Pe\: \D S \right]}.
\end{align*}
With the help of the constraint equations we find
\begin{eqnarray*}
  Z_\rho &=&\e^{-\beta U_2[\E_p,\sigma_p]}  \\
 &\times& \int\DD\vect{e}\ \DD s\;\delta(\div \vect{e})\:\delta(\eps\n\cdot\vect{e}+s)
  \:\mathrm{e}^{-\beta U_2[\vect{e},s]}
\end{eqnarray*}
so that
\begin{equation}
  Z_\rho=\mathrm{e}^{-\beta U_2[\E_p,\sigma_p]}\: \Zfl \label{zfl},
\end{equation}
with $\Zfl$ independent of positions of the mobile charges.  This results in a
complete factorization of the statistical weights of charge configurations
 and of field and surface charge fluctuations
\begin{equation*}
  Z=\biggl[\int\Bigl(\DRj\Bigr)\mathrm{e}^{-\beta U_2[\E_p,\sigma_p]}\biggr]\Zfl
  =\Zcoul \Zfl.
\end{equation*}
We note, however, that following the discussion of Section II, this
fluctuation partition function is not necessarily the true physical
fluctuation partition function that one would calculate for a true physical
metal. Thus sampling with the partition function \Eqref{zrho} will not
necessarily give the correct fluctuation forces acting on electrodes even if,
by construction, it generates the correct relative weight for configurations
of free charges.

\subsubsection*{Algorithm}
The result \Eqref{zfl} is the analytical basis for an algorithm that
simulates fixed-potential surfaces. Sampling the partition function requires
updates generating fluctuations of the variables, $\{\R_j\}$, $\E$ and
$\sigma$. We refer the reader to previous work \cite{PRL88,StatPhys,PRE} for
particle updates and bulk field updates.  We note that the demonstration
requires an integral over the surface charge $\sigma$; thus even if charges in
the volume are discrete those on surfaces should be sampled continuously.

\begin{figure}
  \includegraphics[scale=.5]{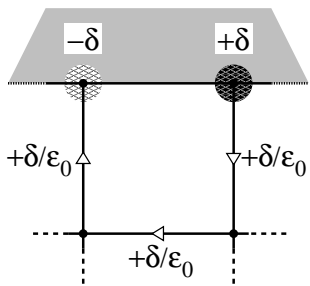}
  \caption{Local update for integrating surface charge fluctuations. Grey
    region represents the conductor volume (where $\E=0$); the white region is
    the dielectric.} \label{surf-local}
\end{figure}

In order to sample the partition function \Eqref{zrho} two new moves need to
be introduced. Firstly, fluctuation in the charge density in the surface of a
given conductor that \textsl{conserve}\/ its total charge. This move is
implemented by a variation of the simplest update for volume charges.  A
random charge amplitude $\delta$ is generated to make a charge pair of
$+\delta$ at one site and $-\delta$ at a second site. Since field lines must
lie outside the conductor, Fig.~\ref{surf-local}, we update the field on three
links connecting the modified sites. To sample the integral \Eqref{zrho} the
charges should be chosen from a continuum distribution.  Secondly, each
conductor $i$ bears a net charge $\int_{S_i}\sigma_i$ which should also
fluctuate.  This is achieved by transferring charges between pairs of
conductors.  Gauss' law then requires modifying the total electric flux
between the two surfaces.

\begin{figure}
  \centering
  \includegraphics{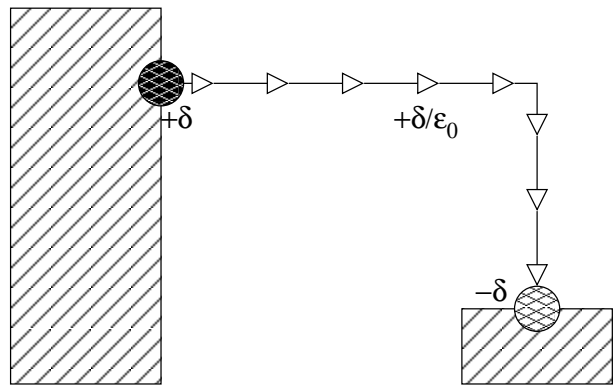}
  \caption{Updating the total charge of conductors implies updating the field
    along a line connecting them. This update is not local.}
\label{surf-global}
\end{figure}

We now demonstrate that despite the need to transfer charge large distances
between conductors this does not dominate the time need to simulate the
system.  We start by noting that  fluctuations on a conductor can be
estimated using an argument based on the capacitance of an isolated conductor
\cite{jancovici}.  In three dimensions the capacitance of an object of size
$d$ scales as $C=\eps d$. If we equate the charging energy to the thermal
energy scale we find
\[
  k_BT= \frac {Q^2} {2 C} \sim \frac {Q^2}{\eps d}
\]
so that the amplitude of charge fluctuation of a conductor is $Q^2 \sim k_B T
\eps d$.  Consider a Monte Carlo trial in which a charge $\delta$ is sent
along a path of length $\ell$, Fig.~\ref{surf-global}. Then \cite{PRE} we
estimate the energy change as $\ell \delta^2/\epsilon_0 a^2$, with $a$ the
mesh spacing.  For this trial to be successful we again require that this
energy is matched to the thermal fluctuations giving $\delta^2 \sim \epsilon_0 a^2 k_B T
/ \ell$. Thus the amplitude of the charge transfer over large distances must
be small.

We now consider $M$ such exchanges. Since charges are transferred in each
direction we require that $M \delta^2 = Q^2$, or $M\sim \ell d/a^2$.  Each
transfer requires a numerical effort that is also $O(\ell/a)$ to update the
links so that a total computational effort ${\eta}=\ell^2 d/a^3$ is needed to
equilibrate the charge fluctuations between the conductors. In any practical
simulation in a box of dimension $L$ we expect that ${\eta}<L^3/a^3$, so
that the effort need to equilibrate charges between conductors is less than
that required to perform a single sweep of the bulk of the simulation.

We thus have the basis for the generalization of a local Monte Carlo algorithm
for simulating charges in the presence of Dirichlet boundary conditions.
Conventional methods for treating this problem use fast Fourier techniques in
simple separable geometries, such as regular simulation cells. The use of a
local formulation of the electrostatics allows one to implement such boundary
conditions in arbitrary geometries --- including rough surfaces or irregular
electrodes where Fourier analysis does not work.

\section{Simulating systems with slab geometries}\label{sec-slab}

We now pass to consideration of a second important class of boundary
conditions that should be imposed in the study of systems with planar
geometries.  Examples include thin quasi two-dimensional slabs isolated in
three-dimensional space, or in the biophysical field simulation of proteins in
a lipid membrane with an implicit solvent. In order to generate the correct
Coulomb interactions between particles with grid based methods the thin sample
must be embedded in a thick three-dimensional simulation box. This box
then contains many more degrees of freedom than the original particle
system. For reasons of efficiency one  wishes to make the repeat distance
perpendicular to the thin sample as small as possible, however if this repeat
distance is too small multiple copies of the sheet ``see'' each other and
introduce artefacts in the simulation.

Imposition of tin-foil or vacuum boundary conditions gives rise to simulations
that converge very slowly as the system size is increased.  The crucial
insight into how to accelerate this convergence was provided in
Ref.~\onlinecite{berkoslab} and efficient implementations are now available that
translate these ideas \cite{holmslab,holmslab2} in molecular dynamics. Our aim
here is to reproduce this rapid convergence in a local formulation of
electrostatics, and to argue that the asymptotic complexity is at least as
good as $N^{3/2}$ and in many systems $N^{1}$.

In standard approaches to the electrostatic energy all the complexity comes
from the transformation from an ambiguous, conditionally convergent sum (a sum
whose answer depends on the order of evaluation) into a rapidly converging,
unambiguous expression, such as the Ewald formula \cite{deleeuw}.  In local
formulations there is \textsl{no equivalent to the conditional convergence}.
Instead one is left with different, inequivalent, choices for the treatment of
the zero wavevector component of the electric field $\E(q =0)$ \cite{JCP120}. Let
us first review the origin of these different choices.

\subsection*{Periodic boundary conditions in local algorithms}

In periodic boundary conditions an arbitrary vector field can be decomposed
into three terms
\[
\E = - \grad \phi + \curl \G + \bar \E
\]
where the potential $\phi$ is periodic, as is $\G$. $\bar \E$ is a constant;
it is the zero wavevector component of the electric field, $\E(q=0)$.  With
this decomposition the energy is given by the sum of three independent terms
\[
U= \frac {\epsilon_0 }{2} \left\{ \int (\grad \phi)^2 \DR + \int (\curl \G)^2
  \DR + V\: \bar \E^2 \right \}
\]
with $V$ the volume of the simulation cell.  Cross-terms are shown to be zero
on integration by parts.  The simplest version of the local Monte Carlo
algorithm uses two updates \cite{PRL88}.
\begin{itemize}
\item Motion of charges $q$ along links $l$ together with a slaved update of
  the electric field on the link according to
  \begin{math}
    E_l \rightarrow E_l - q/ \epsilon_0
  \end{math}.  Both the longitudinal and transverse components of the field
  are modified, as well as the $q=0$ component, $\bar \E$.
\item A plaquette update which leaves unchanged both $\bar \E$ and the
  longitudinal field $-\grad \phi$, modifying only $\G$.
\end{itemize}
This pair of updates are respectively a discretized analog of the two terms on
the right hand side of the Maxwell equation
\[
\epsilon_0 \frac {\partial \E}{\partial t} = -\vect{J} + \curl \vect{H}.
\]
Integrating the Maxwell equation over both time and the simulation volume we
see that the $q=0$ component of the field and the dipole moment of the
system of charges $\vect{d}= \int\!\!\D t \int\!\!\DR \vect{J}$ are linked by
$\epsilon_0 V \bar \E + \vect{d}= \mathrm{const}$, a conservation law which remains
valid for the discretized equations.

Thus the simplest form of the algorithm samples the partition function
\begin{equation}
  Z = \int\DD \E \;\delta (\div \E - \rho) \: \delta(\epsilon_0 V \bar \E + \vect{d})
  \: e^{-\beta \int  \epsilon_0 \E^2/2 \DR }. \label{nowind}
\end{equation}
If we introduce the solution to the Poisson equation in periodic boundary
conditions $\phi_p$ then we see that this partition function samples
configurations with the effective energy
\begin{equation}
  U_\mathit{eff} = \int   \frac{\epsilon_0 (\grad \phi_p)^2}{2} \DR + \frac{\vect{d}^2}
  {2 \epsilon_0 V}. \label{ueff}
\end{equation}
This result is very similar to that found using careful evaluation of the
Coulomb sum \cite{deleeuw} which gives
\[
U_{lps} =\int \frac {\epsilon_0 (\grad \phi_p)^2 }{2} \DR + \frac{\p^2
}{2(1+ 2 \epsilon_s) \epsilon_0 V}
\]
where $\p$ is the cell dipole moment and $\epsilon_s$ the relative
dielectric constant of an ``exterior'' reference medium, two common choices
being $\epsilon_s=1$ for \textsl{vacuum}\/ boundary conditions and
$\epsilon_s=\infty$ for metallic or \textsl{tin-foil}\/ boundary conditions. The
expression resulting from the local algorithm is very similar to that found in
the summation method if we take $\epsilon_s=0$ and we notice that $\vect{d}$ and
$\p$ are \textsl{almost}\/ identical.
$\p$ is the dipole moment of the charge images inside the Bravais cell being
used in the simulation; when a particle crosses the boundary of the cell it is
replaced by that of its periodic images which enters the cell so that $\p$ is
discontinuous. $\vect{d}$ is a dipole moment which is continuous on crossing
Bravais cell boundaries, thus it keeps track of how many times the particles
wind around the simulation box:
\[
\p = \vect{d} - \sum_i \vect{a}_i q_i
\]
where $q_i$ is the charge of each particle in the simulation cell and
$\vect{a}_i$ is a Bravais lattice vector.

The authors of Ref.~\onlinecite{StatPhys} introduced another update scheme for the
electric field based on a \textsl{worm}\/ algorithm widely used in the simulation
of quantum spins \cite{alet}. The algorithm nucleates a pair of virtual
charges $\pm q_v$. One of these charges diffuses until it returns to its
companion where the two charges annihilate. If the mobile charge is confined
to the periodic cell then during the move only the transverse field $\curl \G$
is updated. If the particle is allowed to wind around the cell it breaks the
conservation law on $\epsilon_0 V\bar \E + \vect{d}$, removing the second
$\delta$-function in \Eqref{nowind}. We then find that the effective energy
\Eqref{ueff} is independent of $\vect{d}$; we have effectively periodic, that
is tin-foil, boundary conditions. A similar result is found by including $\bar
\E$ as a single extra dynamic variable which must be updated with a third,
independent Monte Carlo move \cite{PRL88, JCP120b}.

\subsection*{Finite size effects in slab geometries}

We now give a short argument for the slow convergence of the interaction
energy for systems which are simulated in slab geometries. Rigorous
calculations which provide the full justification are to be found in the
literature \cite{fraser, berkoslab,holmslab}. Consider a thin, quasi
two-dimensional system embedded in a simulation box of dimensions $L^2 \times
L_z$; usually one is interested in the case $L_z/L >1$ so that successive
periodic images do not interact too strongly.

\begin{figure}
  \centering
  \includegraphics{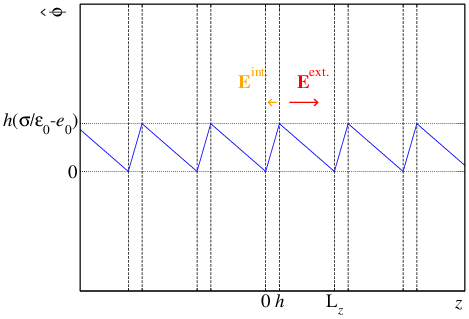}
  \caption{Plot of $\hat \phi(0,z)$ in the presence of a double
    layer. $E^{int}= e_0-\sigma/\epsilon_0$, $E^{ext} = e_0$, with $e_0=\sigma
    h/\eps L_z$. Potential
    plotted over several simulation cells. The non-zero $E^{ext}$ couples
    successive periodic images.}
\label{periodpot}
\end{figure}

The slow convergence of the energy can be understood by considering the
electrostatic potential in a mixed real-space/Fourier representation, $\hat
\phi(\q_\parallel,z)$ where $\q_\parallel$ denotes the transverse
wavevector in the $(x,y)$ plane. Using the fact that polarization of a sample
is equivalent to a space charge $\rho=-\div \P$ 
the mean 
polarization of the thin
sample in the $z$ direction is equivalent to two charge sheets of strength
$\sigma= \pm p_z/L^2 h$ where $p_z$ is the $z$ component of the dipole moment
and $h$ the sample thickness. The interesting physics occurs in the equation
for $q_\parallel=0$. We consider a single cell in the $z$ direction, placing
the source $-\sigma$ at $z=0$ and the second $+ \sigma$ at $z=h$. Then
\[
  \epsilon_0 \frac{\partial ^2 \hat \phi(0, z) }{\partial z^2} = \sigma
  \left [ \delta(z) - \delta(z-h) \right ].
\]
This has as a solution
\begin{align*}
  \hat \phi(0,z) &=(-e_0+\sigma/\epsilon_0 )z  &0&<z<h  \\
  &=(-e_0+\sigma/\epsilon_0 )h -e_0(z-h) &h&<z<L_z
\end{align*}
where we have used continuity, but not periodicity, of the potential.
$e_0$ is
a yet to be determined integration constant; it corresponds to the electric
field flowing between two copies of the simulation cell, Fig.~\ref{periodpot}.
The $z$ component of the electric field is
\begin{align*}
  E_z&= E^{int}=e_0- \sigma/\epsilon_0 \quad &0&<z<h \nonumber \\
  &=E^{ext}=e_0 \quad &h&<z<L.
\end{align*}
The energy of this simulation cell is the integral of $\epsilon_0 E_z^2/2$,
\[
  U_\mathit{Coulomb} = \frac {\epsilon_0 L^2 }{2}
  \left \{
    (e_0-\sigma/\epsilon_0)^2 h + (L_z-h)e_0^2
  \right\}.
\]
The classical tin-foil solution is recovered by setting $e_0= \sigma
h/\epsilon_0 L_z$ so that the potential is periodic, $\phi(0)=\phi(L_z)$,
Fig.~\ref{periodpot}, and
the average field $\bar E_z=\frac{1}{L_z}\int \D z\, E_z(z)=0$. But the energy
then depends on $L_z$,
\begin{equation}
  U_\mathit{Coulomb} = U_\infty - \frac {p_z^2 }{2 \epsilon_0 L^2 L_z} \label{Ustack}
\end{equation}
where $U_\infty$ is the energy in the limit of large $L_z$. If one
considers a series of slabs with identical dipole moment and transverse
dimensions and vary $L_z$ then the energy converges as
only $1/L_z$; very large, and mostly empty boxes must be simulated in order to
reduce finite size artefacts.

\begin{figure}
\centering
\includegraphics{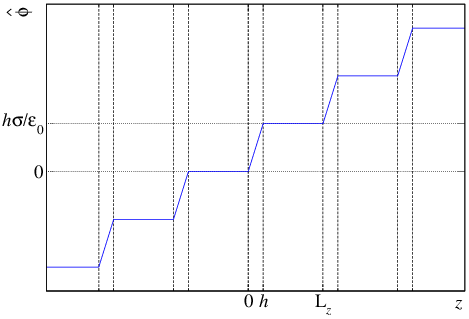}
\caption{Plot of $\hat \phi$ with $e_0=0$. The field between
  slabs, $E^{ext}$, is zero.
 There is no dipolar interaction between the slabs.}\label{stack1}
\end{figure}

There is no such convergence problem when using anisotropic electrostatic
boundary conditions in which the exterior field is always zero: $e_0=0$,
Fig.~\ref{stack1}. The energy is $U_\mathit{Coulomb} = U_\infty$, independent of
$L_z$. The local algorithm automatically gives $e_0=0$, as do Maxwell's
equations; in order to still have $\bar E_x=\bar E_y=0$ we require tin-foil
boundary conditions in the $(x,y)$ plane. Our method implicitly includes the
$+p_z^2/2\eps V$ correction that potential-based algorithms need to add to the
tin-foil energy. Since the particles remain confined to a single lattice cell
in the $z$ direction there is no distinction between $d_z$ and $p_z$.

Let us now consider the effect of density fluctuations with the plane, by
studying the Poisson equation for wavevectors $\q_\parallel \ne 0$:
\[
  -q\lo{}^2\hat{\phi}+\frac{\partial^2\hat{\phi}}{\partial z^2}=
-\frac{{\hat \rho(\q_\parallel,z)}}{\eps}.
\]
Outside of the slab where $\hat \rho=0$ we conclude that
\[
  \hat{\phi}(\q_\parallel\ne 0, z>h) \sim \e^{ \pm q_\parallel  z}
\]
where $\q_\parallel=0$ corresponds to the special case treated above. The
next longest-ranged component to the interaction must come from modes of the
form $( 2 \pi /L,0,0) $ and lead to interactions decaying as $\e^{-2\pi
  L_z/L}$. Already when $L_z/L=3$ one finds a reduction in the interaction by
a factor $7\times 10^{-9}$.

\subsection*{Numerical tests}
To test these ideas we simulated a lattice gas with a slab geometry using the
local algorithm. The lattice spacing $a$ is set to unity. We start a
simulation with all positive and negative charges superposed at $z=0$ and
$\E=0$ in the whole simulation volume. We then displaced positive charges to
$z=1$, creating a dipolar sheet, updating the fields according to the
constrained algorithm.  We simulated the fields using several different
temperatures, $T$ and cell heights, $L_z$.  The total energy can be decomposed
into a static, electrostatic contribution plus a thermal energy $N_{pl} k_B
T/2$ coming from equipartition in the $N_{pl}$ degrees of freedom associated
with the transverse excitations. On fitting the total energy to the
form $$\langle U\rangle =U_\mathit{Coulomb}(L_z) +\frac{ k_BT N_{pl}}{2}$$ we
extracted an estimate of the electrostatic energy as a function of the cell
dimensions.

\begin{figure}
  \centering
  \includegraphics{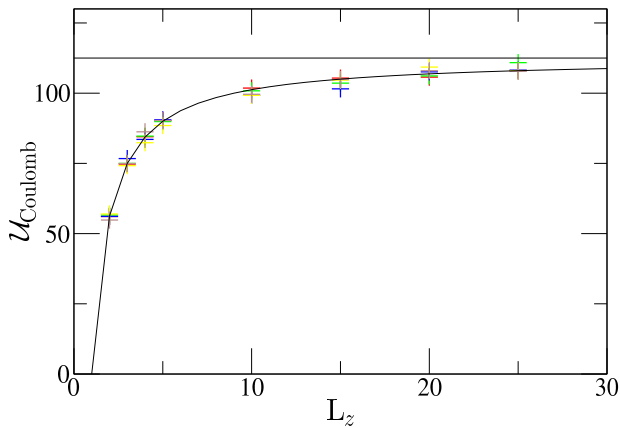}
  \caption{Electrostatic energy of the slab system when all
    three components of $\bar \E$ are sampled independently, corresponding to
    tin-foil boundary conditions. Simulation results (+) follow
    \Eqref{Ustack}.  $L=15$, $\eps=1$. Five values of $k_BT$ from $0.1$ to
    $0.5$ for each $L_z$.}\label{slab00wind}
\end{figure}

\begin{figure}
  \centering
  \includegraphics{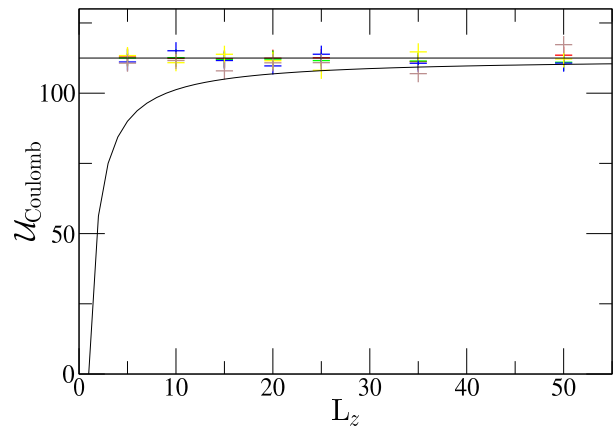}
  \caption{Coulomb energy of the slab system when $\bar{E}_z$
    is not updated independently of charges. Simulation results (+) are
    independent of $L_z$. $L=15$, $\eps=1$. Five values of $k_BT$ from $0.1$
    to $0.5$ for each $L_z$.}\label{slab00}
\end{figure}

In a first series of simulations we imposed the equivalent of tin-foil
boundary conditions by including the three components of $\bar \E$ in the set
of dynamic fields.  The results are shown in Fig.~\ref{slab00wind}; as a guide
to the eye we added the continuous curve which corresponds to a correction in
the energy varying as $1/L_z$ (with the analytically determined prefactor). In
a second series of simulations updates to $\bar{E}_z$ are
dropped. Fig.~\ref{slab00} shows that the electrostatic energy is now
independent of $L_z$ to within the precision of the measurements and equal to
$U_{\infty}$. The natural coupling between $\vect{d}$ and $\bE$ introduced by
our local method removes the dipolar interaction between replicas of the slab.

\subsection*{Discussion}
In a local Monte Carlo simulation the most obvious implementation would use a
mixture of local updates of the particles together with the \textsl{worm}\/
algorithm \cite{StatPhys, alet} in order to integrate over the transverse
degrees of freedom in the empty bulk of the simulation. This worm algorithm
re-equilibrates the transverse field with an effort which is $O(1)$ per
updated degree of freedom (which corresponds to plaquettes). If one were to use
just a single worm update per sweep of the particles then we find a complexity
per sweep which varies as $O(N)$ for the particle motion, and $O(N^{3/2})$ for
the worm dynamics.  In practice launching a worm which simulates the entire
simulation box for every particle sweep results in an oversampling of the
uninteresting transverse degrees of freedom. If one simulates a set of
particles for a time $\tau$ then diffusion motion in the plane will give a
typical displacement $\ell_t\sim \sqrt{\tau}$.  A worm which updates
plaquettes a distance more than $\ell_t$ from the slab is doing unnecessary
work.

This suggests the following hierarchical scheme: every sweep we launch a worm
confined to the simulation slab thickness $h$, then every $4$ sweeps we launch
a worm confined to a distance $2h$. Similarly every $2^{2n}$ sweeps we allow
the worm to propagate $2^nh$ in the $z$ direction. In such a scheme the
longest wavelength modes of the electric field are re-equilibrated on the same
timescale as is needed for particles to diffuse across the simulation
box. The worm then spends most of its time updating at the scale of the slab
so that the total algorithm has a complexity scaling as $N^1$.

\section{Conclusion}

We have considered the formulation of generalized boundary conditions in a
form useful for local electrostatic simulation algorithms. We have shown how
to treat metallic boundaries for which one imposes a constant potential at a
surface. An additional Monte Carlo move which transfers charge between
conductors performs the Legendre transformation from a constant-charge to a
constant-potential ensemble.

The local algorithm (like Maxwell's equations) naturally generates a dipolar
contribution to the solution to the Poisson equation in a periodic simulation
cell. By choosing an anisotropic integration over the $q=0$ components of the
electric field we reduce the spurious interaction between different copies of
a planar system.

Molecular dynamics implementations of both problems are potentially possible
\cite{joerglong}.  For the first problem of constant-potential simulation one
could alternate between a symplectic integrator (e.g. velocity Verlet) for the
particles and a Monte Carlo step for charge transfer between conductors, or
introduce a kinetic degree of freedom for the charges on each conductor. In
the second problem of slab geometries one would need to follow the evolution
of the electric field at each time step throughout the simulation cell,
loosing the possibility of multiscale updates.
Implementation of these ideas has been performed in Ref.~\onlinecite{comp}.

Work financed in part by Volkswagenstiftung.
\bibliography{coulomb}
\end{document}